\documentclass[12pt,a4paper,oneside]{article}

\usepackage[top=20truemm,bottom=20truemm,left=20truemm,right=20truemm]{geometry}

\usepackage[dvipdfmx]{hyperref}
\usepackage{amsmath, amsthm}
\usepackage{amssymb}
\usepackage{multirow}
\usepackage{fancybox}
\newcommand{\Trp}{\text{Tr}'}
\numberwithin{equation}{section}

\begin{document}
\begin{flushright}
\end{flushright}
\begin{center}
{\LARGE Localization of four-dimensional super Yang-Mills theories compactified on Riemann surface}\\
\vspace{5mm}
{\large Koichi Nagasaki${}^1$}\\
\vspace{5mm}
{\small ${}^1$Department of Physics and Center for High Energy Physics,
Chung Yuan Christian University\\
Address: 200 Chung Pei Road, Chung Li District, Taoyuan City, Taiwan 32023\\
mail: {koichi.nagasaki24@gmail.com}}\\
\end{center}
\vspace{3cm}

\abstract{
We consider the partition function of super Yang-Mills theories defined on $\mathbb{T}^2 \times \Sigma_\bold{g}$. 
This path integral can be computed by the localization. 
The one-loop determinant is evaluated by the elliptic genus.
This elliptic genus gives trivial result in our calculation.
As a result, we obtain a theory defined on the Riemann surface.}

\section{Introduction and Summary}
Studying theories which are compactified on some lower dimensional manifolds often has a very interesting results. 
AGT correspondence \cite{Alday:2009aq, Wyllard:2009hg} is an example of such theories obtained by compactification of six-dimensional theory on Riemann surfaces.
In the past works \cite{Nagasaki:2014xya, Nagasaki:2015xsa}, we considered super Yang-Mills theories (SYM) defined on flat spacetime times the Riemann surface.
This theory is twisted to preserve the supersymmetry on the curved space.
In the case where the Riemann surface has a boundary we also consider the boundary condition for preserving the supersymmetry.
So the localization method is valid for our case.

The localization method is a very useful technique to calculate the path integral for theories with supersymmetries. 
This technique has been studied for theories defined on various spaces \cite{Pestun:2007rz, Nekrasov:2002qd, Hama:2011ea, Imamura:2011wg, Nagasaki:2011sh, Nawata:2011un, Kawano:2012up, Fukuda:2012jr, Fujitsuka:2013fga, Lee:2013ida, Hosomichi:2015jta, Kawano:2015ssa, Honda:2015yha}.

In this paper we would like to compute the partition function of such SYM defined on the product space of Riemann surfaces and flat space.
To obtain the finite result we consider in this paper the Euclidean theory on compact space $\mathbb{T}^2\times \Sigma_\bold{g}$.
This partition function is defined by path integral:
\begin{align}
Z_{\mathbb{T}^2\times \Sigma_g} = \int [d\Phi] \text{e}^{-\int d^4x\mathcal{L}[\Phi]}.
\end{align}
The partition function on the torus is evaluated by an object called ``elliptic genus.''
This quantities have been calculated in some examples \cite{Benini:2013xpa, Benini:2013nda}.

In this paper we treat two-dimensional superfields introduced in the previous paper \cite{Nagasaki:2015xsa} by following the method of \cite{Erdmenger:2002ex} which treat the three-dimensional superfields to construct four-dimensional theory with a boundary.
In our case we introduced the boundary on the curved space. 
Then the remaining symmetry is smaller than the case in \cite{Erdmenger:2002ex} which treats the boundary on the flat space.
In our case the remaining symmetry is two-dimensional $\mathcal{N}=(2,2)$.
We define the supersymmetry transformation to that these fields preserve the Wess-Zumino gauge. 
Using this supersymmetry we calculate the partition function by the localization method. 
In this paper the calculation is performed on the closed Riemann surface (no boundary) for simplicity.

For future works considering theories with boundaries is interesting.
A geodesic boundary can be introduced so that the theory preserves the supersymmetry. 
The condition for preserving half of the supersymmetries on a closed Riemann surface is studied with various numbers of the supersymmetry in \cite{Nagasaki:2014xya}.
The $\mathcal{N}$ = (2,2) supersymmetry considered here is broken in the case of the boundary theory to half of it, namely, $\mathcal{N}$ = (1,1).
The boundary conditions for various manifold is considered in \cite{1995NuPhB.455..522M, Hori:2000ck, Petkova:2000dv, Hori:2000ic, Vassilevich:2003xk, Cardy:2004hm, Herbst:2008jq, Berman:2009kj, Berman:2009xd, Okazaki:2013kaa}.
Localization on curved space with boundary is recently studied. For example, the partition functions and the Wilson loops are studied by localization on three-sphere and two-sphere case \cite{Sugishita:2013jca}.

\subsection*{Main result}\label{Sec:SummaryDisc}
In this paper the partition function of SYM on $\mathbb{T}^{2}\times\Sigma_\bold{g}$ is calculated.
The integral on the torus is performed and the theory on the Riemann surface is obtained:
\begin{align}\label{Eq:Mainresult}
Z_{\mathbb{T}^2\times \Sigma_\bold{g}} 
&= \prod_{x^2,x^3\in \Sigma_\bold{g}} \int du \int_{\mathcal{M}} \mathcal{D}\Phi_0.
\end{align}
where $u$ is the holonomy which is related to the components of the gauge field on the torus and moduli parameter as:
\begin{align}\label{Eq:defHolonomy}
u = \oint A_t dt - \tau\oint A_s ds, 
\end{align}
where $A_t$ and $A_s$ are the temporal and spatial components of the gauge field.
The second integral in eq.\eqref{Eq:Mainresult} is performed over the moduli space which is defined by the fixed point of the localization and $\mathcal{D}\Phi_0$ is the measure on it.
The integrand is the elliptic genus but it is equal to one in our result, that is, the one loop determinant is trivial.
This is a similar situation to \cite{Lee:2013ida}, where SYM defined on $S^2\times $ (3-manifold) is considered. 
Then, we see a similar reduction from 4d-theory defined on the product space to 2d theory on the Riemann surface:
\begin{align}
Z_\text{4d,SYM} [\mathbb{T}^2\times \Sigma_{\bf g}] = Z_\text{2d,CFT}[\Sigma_{\bf g}].
\end{align}

On general surfaces with arbitrary genus $\bold{g}$ calculating the above integral seems to be very hard.
For future work, I would like to calculate them in some simple cases, which is sphere $S^2$ for example, and will reveal the dependence of the theory on the geometry of the surface.

This paper is organized as follows:
In section \ref{Sec:2dsuperfields} we define the superfields defined on the two-dimensional space. 
This is introduced in \cite{Nagasaki:2015xsa} but we now redefined it in the Euclidean case.
We confirm this theory surely has 2d $\mathcal{N}=2$ supersymmetry. 
In section \ref{Sec:ClassicalPart} using this supersymmetry, we identify the fixed points of localization.
Evaluating the integrand at this point, the classical part of the partition function is obtained.
The one-loop determinant is evaluated in section \ref{Sec:Oneloopdet}. 
We investigate the chiral multiplet part and the vector multiplet part separately. 
Finally we see that the resulting partition function represents the theory localized on the Riemann surface.
Appendix \ref{Sec:EtaTheta} gives the definition of eta and theta functions used in the paper.

\section{Construction by the two-dimensional superfields}\label{Sec:2dsuperfields}
\subsection{Setup}\label{Sec:Setup}
First we review the setup of the four-dimensional SYM with $\mathcal{N}=4$ supersymmetry. 
We consider SYM defined on $\mathbb{T}^{2}\times \Sigma_\bold{g}$, where $\mathbb{T}^{2}$ is the torus with complex structure $\tau$ and $\Sigma_\bold{g}$ is the Riemann surface with genus $\bold{g}$.
Our theory consists of a four-dimensional gauge field $A_\mu, \mu=0,1,2,3$, six scalars $X^A, A=4,5,\cdots, 9$ and the Majorana-Weyl spinor $\Psi$ which satisfies $\Gamma_{0123456789} \Psi = \Psi$.

To preserve some of supersymmetries on a curved manifold we twist the theory \cite{Nagasaki:2014xya}.
The twisted derivative is defined by the external gauge field $\mathcal{A}_{\mu}$ in SO(2)$^3$ $\subset$ SO(6) R-symmetry as
\begin{subequations}
\begin{align}
D_\mu \Phi_A &= \partial_\mu \Phi_A + i[A_\mu, \Phi_A] + \sum_{B}\mathcal{A}^{AB}_\mu \Phi_B,\\
D_\mu \Psi &= \partial_\mu \Psi + i[A_\mu, \Psi] + \frac14 \Omega^{AB}_\mu \Gamma_{AB}
			- i\mathcal{A}_\mu \Psi.
\end{align}
\end{subequations}
$\mathcal{A}_\mu$ and $\mathcal{A}_{\mu}^{AB}$ are related by the Gamma matrices: $\mathcal{A}_\mu = \frac12 \mathcal{A}_\mu^{AB}\Gamma_{AB}$, where $(A,B)$ takes values $(4,5)$, $(6,7)$, or $(8,9)$.  

The Minkowski action is given in \cite{Nagasaki:2015xsa}.
In order to calculate the path integral we use the Euclidean one;
\begin{align}\label{Eq:YMlagrangian}
\mathcal{L}_\text{gauge} = \frac{1}{(g_\text{YM})^2}\sqrt{g}\:\text{Tr}'
		\Big\{
			 + \frac{1}{4}F_{\mu\nu}F^{\mu\nu}
			+ \frac{1}{2}D_\mu X_A D^\mu X^A
			- \frac{1}{4}[X_A,X_B][X^A,X^B]
			+ \frac{R}{4} \sum_{A=4,5}(X_AX_A)\nonumber\\
			- \frac{1}{2}\overline{\Psi}\Gamma^\mu D_\mu \Psi
			+ \frac{1}{2}\overline{\Psi}\Gamma^A[X_A,\Psi]
			\Big\}.
\end{align}
We use the $\text{Tr}'$ normalized as $\text{Tr}' = \frac{1}{h^{\vee}}\text{Tr}_\text{ad}$ where $h^{\vee}$ is the dual Coxeter number.
In the mass term, $A = 4,5$ is singled out because the directions $4,5$ is related to the twisting.
It is the same action used in \cite{Nagasaki:2014xya}.

\subsection{Action by 2d superfields}

In this section we construct the Euclidean action \eqref{Eq:YMlagrangian} in terms of two-dimensional superfields defined on $\mathbb{T}^2$.
First we define the coordinates on the spacetime. 
Let $(x^0,x^1)$ be coordinates on the torus and $(x^2,x^3)$ be those on the Riemann surface.
The metric is 
\begin{align}
g_{mn} = \delta_{mn},\;\;
g_{ij} = e^{2h(x^2,x^3)}\delta_{ij}.
\end{align}
The indices $m, n$  are used for the coordinates on the flat directions (on the torus) and the indices $i,j$ are used for the curved space (the Riemann surface). $h(x^2,x^3)$ is the function so that it has constant curvature on the Riemann surface $R=+2,0,-2$ for $\bold{g}=0$, $\bold{g}=1$, $\bold{g}\geq 2$, respectively.
We use the sigma matrices in unsusual ordered as
\begin{align}
\sigma^{M} =
	\left\{
		\sigma^{m}, \sigma^{a}
	\right\}
	= \left\{
\left(\begin{array}{cc}-1 & 0 \\0 & -1\end{array}\right),
\left(\begin{array}{cc}1 & 0 \\0 & -1\end{array}\right),
\left(\begin{array}{cc}0 & 1 \\1 & 0\end{array}\right),
\left(\begin{array}{cc}0 & -i \\i & 0\end{array}\right)
\right\}.
\end{align}
We use the indices $a,b$ to label the coordinates on the dimensional reduced directions: 6 and 8. 
It is more convenient to use this for this work.
We denote $M$ to summarize the coordinates on the torus and these space: $M=0,1,6,8$.

We introduce two kinds of superfields --- vector and chiral superfields.
We need one vector superfield and three chiral superfields to realize the degrees of freedom of the 4d SYM.
The vector superfield can be expanded as  
\begin{align}
V 	&= -\sum_{m=0,1}\theta\sigma^m \bar\theta v_m (x)
	 -\sum_{a=6,8}\theta\sigma^a \bar\theta v_a (x)
	 + i\theta\theta\bar\theta\bar\lambda(x)
	 - i\bar\theta\bar\theta\theta\lambda(x)
	 + \frac{1}{2}\theta\theta\bar\theta\bar\theta D(x).
\end{align}
We also use the linear multiplet defined by the vector superfield:
\begin{align}
\Sigma &:= \frac{1}{2\sqrt{2}}\big\{\overline{\mathcal{D}}_{+},\mathcal{D}_{-}\big\}.
\end{align}
We used the supercovariant derivatives:
\begin{align}
\mathcal{D}_{\alpha} = \text{e}^{-V} D_\alpha \text{e}^{V},\:\:
\overline{\mathcal{D}}_{\alpha} = \text{e}^{V} \overline{D}_\alpha \text{e}^{-V},
\end{align}
where
\begin{align}
D_{\alpha} := \frac{\partial}{\partial \theta^{\alpha}} 
	+ i(\sigma^\mu \bar\theta)_{\alpha}\frac{\partial}{\partial x^\mu},\:\:
\overline{D}_{\dot\alpha} := -\frac{\partial}{\partial \theta^{\dot\alpha}} 
	- i(\theta\sigma^\mu )_{\dot\alpha}\frac{\partial}{\partial x^\mu}.
\end{align}
Supersymmetry charge are defined similarly:
\begin{align}
Q_{\alpha} := \frac{\partial}{\partial \theta^{\alpha}} 
	- i(\sigma^\mu \bar\theta)_{\alpha}\frac{\partial}{\partial x^\mu},\:\:
\overline{Q}_{\dot\alpha} := -\frac{\partial}{\partial \theta^{\dot\alpha}} 
	+ i(\theta\sigma^\mu )_{\dot\alpha}\frac{\partial}{\partial x^\mu}.
\end{align}

The three chiral multiplet has the following expansion:
\begin{align}
\Phi_i = \phi_i + \sqrt{2}\theta\psi_i + \theta\theta F_i
	+ i\theta\sigma^m \bar\theta \partial_m \phi_i
	+ \frac{i}{\sqrt{2}}\theta\theta\bar\theta\bar\sigma^m \partial_m\psi_i
	+ \frac14 \theta\theta\bar\theta\bar\theta \square\phi_i.
\end{align}
The gauge transformation of the vector superfield is defined in the usual way:
\begin{align}\label{Eq:GaugetrV}
\text{e}^{2V} &\rightarrow \text{e}^{-i\Lambda^{\dagger}}\text{e}^{2V}\text{e}^{i\Lambda},
\end{align}
while that of chiral superfields is
\begin{subequations}\label{Eq:GaugetrCh}
\begin{align}
\Phi_1 &\rightarrow \text{e}^{-i\Lambda}\Phi_1 \text{e}^{i\Lambda},\\
\Phi_i &\rightarrow \text{e}^{-i\Lambda}\Phi_i \text{e}^{i\Lambda}
	 - \text{e}^{-i\Lambda}\frac{\partial_i}{2} \text{e}^{i\Lambda}, \;\; i= 2,3\label{Eq:GaugetrChiral23},
\end{align}
\end{subequations}
where $\Lambda$ is a chiral superfield.

The relation between these component fields and 4d SYM fields is shown in Table \ref{Table:CorrespondenceFields}.
In this table the projection $P_{1\pm}$, $P_{2\pm}$ decomposes 16-components of the fermion $\Psi$ into four 4-component Weyl-spinors.

The Lagrangian by superfield formalism is given in \cite{Nagasaki:2015xsa}.
Now we use the Euclidean action:
\begin{align}\label{Eq:YMactionSF}
\mathcal{L}[V,\Phi_i]
		&= - \sqrt{g}\:{\Trp}\int d^4\theta \bigg[ -\overline{\Sigma}\Sigma 
			+ 2e^{-2V}\overline{\Phi}_1e^{2V}\Phi_1
			\bigg]
			- {\Trp}\int d^4\theta \sum_{i=2,3} \bigg\{e^{-2V}\left(\frac12\partial_i +\overline{\Phi}_i\right)e^{2V} + \Phi_i\bigg\}^2
				\nonumber\\
		&\hspace{1cm}	-  2{\Trp} \bigg[ 
		 \int d^2\theta\: \Phi_1(\partial_2 \Phi_3 - \partial_3 \Phi_2
					-2[\Phi_2,\Phi_3] )+ \text{  c.c.  }
				\bigg].
\end{align}
As shown in \cite{Nagasaki:2015xsa} this action gives the 4d SYM action \eqref{Eq:YMlagrangian}.
\begin{table}[b]
\centering
  \begin{tabular}{| l  c  l | l |}
    \hline
    2d fields 		&  				& 4d fields & relation\\ \hline
    $v_0$, $v_1$ 	& $\longleftrightarrow$ & $A_0$, $A_1$ & $v_0 = A_0,\:\: v_1 = A_1$ \\ \hline
    $v_6$, $v_8$ 	& $\longleftrightarrow$ & $X_6$, $X_8$ & $v_6 = X_6,\:\: v_8 = X_8$ \\ \hline
    $\:\: \lambda$ 	& $\longleftrightarrow$ & $P_{1-}P_{2+}\Psi$ &  \\ \hline
    $\:\: \phi_1$	& $\longleftrightarrow$ & $X_7$, $X_9$ & $\phi_1 = \frac{1}{2}X_7 + i\frac{1}{2}X_9$ \\ \hline
    $\:\: \psi_1$  & $\longleftrightarrow$ & $P_{1-}P_{2-}\Psi$ & \\ \hline
    $\:\: \phi_2$ 	& $\longleftrightarrow$ & $A_2$, $X_4$ & $\phi_2 = \frac{1}{2}X_4 - i\frac{1}{2}A_2$ \\ \hline
    $\:\: \psi_2$ 	& $\longleftrightarrow$ & $P_{1+}P_{2+}\Psi$ & \\ \hline
    $\:\: \phi_3$ 	& $\longleftrightarrow$ & $A_3$, $X_5$ & $\phi_3 = \frac{1}{2}X_5 - i\frac{1}{2}A_3$ \\ \hline
    $\:\: \psi_3$ 	& $\longleftrightarrow$ & $P_{1+}P_{2-}\Psi$ & \\ \hline
  \end{tabular}
\caption{Correspondence between components of superfields and 4d fields}
\label{Table:CorrespondenceFields}
\end{table}


\subsection{2d $\mathcal{N}=(2,2)$ supersymmetry}
We define supersymmetry transformation of superfields, $\delta_S$, by combining the usual one, $\xi Q + \bar\xi \overline{Q}$, and the gauge transformation \eqref{Eq:GaugetrV} and \eqref{Eq:GaugetrCh} to preserve the Wess-Zumino gauge.
Supersymmetry transformation for vector multiplet is expressed as follows in components:
\begin{subequations}\label{Eq:DeltaSVec}
\begin{align}
\delta_S v_m &= i(\xi\sigma_m \overline{\lambda} + \bar\xi\bar\sigma_m \lambda) 
				-\frac{i}{2} \nabla_m(\Lambda_0 - \Lambda_0^*),\\
\delta_S v_a &= i(\xi\sigma_a \overline{\lambda} + \bar\xi\bar\sigma_a \lambda) 
				-\frac{i}{2}[iv_a, \Lambda_0 - \Lambda_0^*],\\
\delta_S \overline{\lambda}^{\dot\alpha} &= -i\bar\xi^{\dot\alpha} D 
				+ (\bar\sigma^{MN}\bar\xi)^{\dot\alpha} v_{MN}
				- \frac{1}{2}[\Lambda_0 - \Lambda_0^*, \overline{\lambda}^{\dot\alpha}],\\
\delta_S {\lambda}_{\alpha} &= i\xi_{\alpha} D 
				+ (\sigma^{MN}\xi)_{\alpha} v_{MN}
				- \frac{1}{2}[\Lambda_0 - \Lambda_0^*, {\lambda}_{\alpha}],\\
\delta_S D &= \nabla_M ( -\xi\sigma^{M}\overline{\lambda} + \bar\xi\bar\sigma^{M}\lambda)
				- \frac{i}{2}[\Lambda_0 - \Lambda_0^*, {D}],
\end{align}
\end{subequations}
where $\theta$ independent component of the chiral superfield can be set to zero, $\Lambda_0=0$. 
Here we defined the gauge covariant derivative as $\nabla_m  := \partial_m + i[v_m,*]$.


Using the correspondence between superfields and 4d gauge fields (Table \ref{Table:CorrespondenceFields}), 
the SUSY transformation of the chiral superfields are rewritten as follows:\\
For $i=1$,
\begin{subequations}
\begin{align}
\delta_S \phi_1 &= \sqrt{2}\xi\psi_i,\\
\delta_S (\psi_1)_\alpha &= \sqrt{2}\xi_\alpha F_1
	+ \sqrt{2}i (\sigma^m \bar\xi)_{\alpha} \nabla_m \phi_1
	- \sqrt{2}(\sigma^a\bar{\xi})_\alpha [v_a, \phi_1],\\
\delta_S F_1 &= \sqrt{2}i \bar\xi \bar\sigma^m \nabla_m \psi_1
	-\sqrt{2}\bar{\xi}\bar{\sigma}^a [v_a, \psi_1]
	+ 2i\bar\xi [\overline{\lambda}, \phi_1],
\end{align}
\end{subequations}

while for $i=2,3$,
\begin{subequations}
\begin{align}
\delta_S \phi_i &= \sqrt{2}\xi\psi_i,\\
\delta_S (\psi_i)_\alpha &= \sqrt{2}\xi_\alpha F_i
	+ \frac{1}{\sqrt{2}} (\sigma^m \bar\xi)_{\alpha} v_{mi}
	- \frac{1}{\sqrt{2}} (\sigma^a\bar{\xi})_\alpha \nabla_i v_a\nonumber\\
&\qquad
	+ \frac{i}{\sqrt{2}} (\sigma^m \bar{\xi})_\alpha \nabla_m X_{i+2}
	- \frac{1}{\sqrt{2}} (\sigma^\mu \bar{\xi})_\alpha [v_a, X_{i+2}],\\
\delta_S F_i &= \sqrt{2}i \bar\xi \bar\sigma^m \nabla_m \psi_i
	-\sqrt{2} \bar{\xi}\bar{\sigma}^a [v_a, \psi_i]
	+ i\bar\xi [\overline{\lambda}, X_{i+2}]
	+ i\bar{\xi}\nabla_i \overline{\lambda}.
\end{align}
\end{subequations}
Note that there are additional terms in the transformation of $\Phi_2$ and $\Phi_3$ compared to $\Phi_1$ because of the strange gauge transformation for $\Phi_2$ and $\Phi_3$ \eqref{Eq:GaugetrChiral23}. 
This component expression represents gauge invariance manifestly. 
The gauge covariant derivative is defined as $\nabla_M  = \partial_M + i[v_M, *]$, $(M=0,1,6,8)$.
Note that for $a$ directions it is simply a commutator $\nabla_a = [v_a, *]$.

\section{Localization}\label{Sec:ClassicalPart}

Now the Lagrangian is expressed by the exact form since it is the integral form $\int d^2\theta (\cdots)$ as shown in eq.\eqref{Eq:YMactionSF}.
Following the usual method of localization, the path integral is evaluated in the limit $g_\text{YM} \rightarrow 0$.
The integral is localized to $\mathcal{L}_\text{bos}=0$, where $\mathcal{L}_\text{bos}$ is the bosonic part of the Lagrangian.
Then the fixed locus is identified by completing the square:
\begin{align}
\mathcal{L}_\text{bos} 
&= \frac12\Big\{
D -2i\Big(
	\sum_{\ell=1,2,3} [\phi_{\ell},\overline{\phi}_{\ell}]
	- \sum_{i=2,3} \partial_i \text{Re}\phi_i
	\Big)
\Big\}^2
	+ 2\Big(
	\sum_{\ell=1,2,3} [\phi_{\ell},\overline{\phi}_{\ell}]
	- \sum_{i=2,3} \partial_i \text{Re}\phi_i
	\Big)\nonumber\\
&\qquad
	+ \frac12 v_{01}^2 + \frac12\sum_{a=6,8} \nabla^mv_a \nabla_mv_a
	+ |[v_6, v_8]|^2 + 2\sum_{\ell} |F_\ell |^2\nonumber\\
&\qquad
	+ 2\overline{\nabla^m \phi_1} \nabla_m \phi_1
	+ 2\sum_{i=2,3} \nabla^m \text{Re}\phi_i \nabla_m \text{Re}\phi_i
	+ 2\sum_{\ell, a} |[v_a, \overline{\phi}_{\ell}]|^2\nonumber\\
&\qquad
	+ \frac12 \sum_{i=2,3} (\partial_i v^M + 2\nabla^M \text{Im}\phi_i)
		(\partial_i v_M + 2\nabla_M \text{Im}\phi_i).
\end{align}
The fixed point is obtained by demanding that each term is zero:
\begin{align}\label{Eq:Fixedpt}
\mathcal{L}_\text{bos,E}= 0 \Leftrightarrow\nonumber\\
& D=0,\qquad
\sum_{\ell=1,2,3} [\phi_{\ell},\overline{\phi}_{\ell}] - \sum_{i=2,3} \partial_i\text{Re}\phi_i =0,\nonumber\\
& v_{MN} = 0,\qquad
 F_{\ell} = 0, \nonumber\\
& \nabla_m \phi_1 = 0,\qquad
 \nabla_m \text{Re}\phi_i = 0, \qquad
 [v_a, \phi_\ell] = 0, \nonumber\\
& \partial_i v_{m} + 2\nabla_m\text{Im}\phi_i = 0.
\end{align}
In the above the indix $\ell$ runs 1,2 and 3.
The indices on the torus are $m,n$ and those of the Riemann surface are $i,j$.
The indices $a,b$ sun $6,8$ and we use $M,N$ to summarize $m$ and $a$: $M=(m,a) = (0,1,6,8)$.
Note that $v_{MN}$ is the abbreviation for $v_{01}$, $v_{ma}$ and $v_{68}$, that is, $v_{MN}=0$ means
\begin{align}\label{Eq:Fixedptv}
v_{01} = 0,\:\:
v_{ma} = \nabla_mv_a=0,\:\:
[v_6,v_8] = 0.
\end{align}
At the locus satisfying \eqref{Eq:Fixedpt}, the Lagrangian is evaluated as 
Its bosonic part is 
\begin{equation}
{\mathcal{L}}_\text{bos} = 0.
\end{equation}
It gives the classical part of the partition function.
In the below, we calculate the one-loop determinant.

\section{One-loop determinant}\label{Sec:Oneloopdet}
The partition function is factorized into the classcal part which we evaluated in the previous section and the Gaussian integral around the localized locus. It is the one-loop determinant $Z_\text{one-loop}$. 
\begin{align}
Z_{\mathbb{T}^2\times \Sigma_g} 
	= \lim_{g_\text{YM} \rightarrow \infty}
	 \int [\mathcal{D}\Phi] \exp\left[ - \int d^4 x
	  	\mathcal{L}(\Phi) 
		\right]
	= \int [\mathcal{D}\Phi_0] \:Z_\text{cl} (\Phi=\Phi_0) \cdot Z_\text{one-loop}.
\end{align}

The one-loop determinant is evaluated by the quantity which is called ``the elliptic genus ''
\cite{Witten:1993jg, Witten:1993yc, Benini:2013xpa, Benini:2013nda}.
In the previous section we saw the classical Lagrangian vaniches at the fixed point, $\mathcal{L}_\text{cl}=0$.
Then, the classical part of the partition function is trivial: $Z_\text{cl} = 1$.

Let us perform the path integral on the torus to evaluate the one-loop part. 
This qunatity is the elliptic genus.
The elliptic genus is defined as 
\begin{align}
Z = \text{Tr}_{\small\text RR} (-1)^F q^{H_L} \bar{q}^{H_R} y^J \prod_{a} x_a^{K_a},
\end{align}
where RR under the trace means the fermions have periodic boundary conditions, $F$ is the fermion number, $H_L(R)$ is the left (right) moving hamiltonian: $2H_{L,R} = H \pm iP$, $J$ is the left moving R-current and $K_a$ is flavor symmetry group, which is zero in our case.
We defined the following variables:
\begin{align}
q := \text{e}^{2\pi i\tau}, \:
y := \text{e}^{2\pi i z}, \:
x_a := \text{e}^{2\pi i u_a}.
\end{align}
These are related to the complex structure, R-symmetry and flavor-symmetry, respectively. 
Let $A^R$ and $A^\text{flavor}$ be the R-symmetry and flavor-symmetry gauge fields. $z$ and $u_a$ are related to them as
\begin{align}
z = \oint_t A^R - \tau\oint_s A^R, \:
u_a = \oint_t A^\text{flavor} - \tau\oint_s A^\text{flavor}.
\end{align}
We choose the R-charge as $Q[\theta^{-}] = 1$ and $Q[\theta^{+}] = 0$.

We calculate the one-loop part in the free field approximation. The free Lagrangian of the theory is 
\begin{align}
\mathcal{L}[V,\Phi_i]|_\text{free} &=
	\frac12 D^2 + \frac12 v_{01}
	+ \partial^m \sigma \partial^m \sigma
	+ i (\lambda_{+} (\partial_0 - \partial_1) \lambda_{+})
	+ i (\lambda_{-} (\partial_0 + \partial_1) \lambda_{-})
	+ \sum_{i=1,2,3} \overline{F}_iF_i\nonumber\\
&\qquad	+ \sum_{i=1,2,3} 2 \partial^m \phi_i \partial_m \phi_i
	+ 	\sum_{i=1,2,3} i\psi_i \sigma^m \partial_m \psi_i.
\end{align}

\subsection{Chiral multiplet part}\label{Subsec:chiralmult}
We introduced three chiral multiplets.
In the following expression we denote the U(1) charge of the chiral multiplet as $\alpha$.
The chiral multiplet can be expanded by the generator of Cartan subalgebra and roots:
\begin{align}
\Phi_i = {\Phi_i}_a H^a + {\Phi_i}_\alpha E^\alpha.
\end{align}
where $a$ runs from $1$ to $N-1$ and $\alpha$ runs $1$ to $N^2-N$ for gauge group SU($N$).
By considering each components, for each root $\alpha$ the partition function is roughly,
\begin{align}
Z_\text{chiral}(\alpha) 
\sim y^{\alpha - \frac12} 
	\frac{1- y^{-\Delta +1}}{1- y^{\Delta}}
	\prod_{n=1}^{\infty}
	\frac{1-q^n y^{-\Delta+1}x^\alpha}{1-q^n y^\Delta x^\alpha}
	\frac{1-q^n y^{\Delta-1}x^\alpha}{1-q^n y^{-\Delta}x^\alpha}
= \frac{\theta_1(y^{-\Delta +1}x^{\alpha},q)}{\theta_1(y^{\Delta}x^{\alpha},q)},
\end{align}
where $\Delta$ is the R-charge of the chiral multiplet.
In the last expression we used theta function summarized in Appendix \ref{Sec:EtaTheta}.

The Cartan part, $\alpha=0$, can be calculated in the similar way except for the zero mode, $(\text{the powers of } y) =0$.
This mode gives divergent. 
Then, we need to remove this mode to get a finite result by removing the mode corresponding $n=0$.
By the relation between theta and eta functions,
\begin{align}
\lim_{\xi\rightarrow 1}\frac{1}{\xi-1} \theta_1(\xi,q) = i\eta(q)^3.
\end{align}
Removing the zero mode corresponds to replacing $\theta_1$ in the numerator of the above partition function by $\eta$.

Then, the elliptic genus for $\Phi_1$ ($\Delta=1$) is 
\begin{align}\label{Eq:Cartzero}
Z_{\Phi_1} 
= \left(\frac{i\eta(q)^3}{\theta_1(y,q)}\right)^{N-1}
 	 \prod_{\alpha}\frac{\theta_1(x^\alpha,q)}{\theta_1(y x^\alpha,q)}.
\end{align}
The elliptic genus for $\Phi_2$ and $\Phi_3$ ($\Delta=0$) is 
\begin{align}\label{Eq:Cartzero}
Z_{\Phi_{i}} 
= \left(\frac{\theta_1(y^{-1},q)}{i\eta(q)^3}\right)^{N-1}
 	 \prod_{\alpha}\frac{\theta_1(y^{-1} x^\alpha,q)}{\theta_1(x^\alpha,q)}
	 \qquad (i=2,3).
\end{align}

\subsection{Vector multiplet part}\label{Subsec:Vectormult}

Vector multiplet has the components $(v_{-}, v_{+}, \sigma, \lambda_{-}, \lambda_{+}, D)$.
The charges of these fields are $0,0,-1,0,+1$ and $0$, respectively.
The partition function of vector multiplet decomposes as
\begin{equation}
Z_\text{vec} = (Z_{\text{vec;cart}})^{r} \times Z_{\text{vec;off}} ,
\end{equation}
where the first factor is the contribution from the Cartan part of the Lie algebra and the second factor is the contribution form the off-diagonal part. The powers of $r$ is derived from the rank of the gauge group; for example, $N-1$ for SU($N$).

The off-diagonal part consists of a product over root vectors:
\begin{align}
Z_\text{vec;off}  = \prod_{\alpha;\text{root}} \frac{\theta_1(x^{\alpha}, q)}{\theta_1(y^{-1}x^{\alpha}, q)}.
\end{align}
The Cartan part is obtained by removing zero modes in the same way as chiral mutliplets:
\begin{align}
Z_\text{vec;cart} = \left(\frac{i\eta(q)^3}{\theta_1(y^{-1},q)}\right)^{N-1}.
\end{align}
Then the contribution from the vector multiplet is
\begin{align}
Z_\text{vec} = \left(\frac{i\eta(q)^3}{\theta_1(y^{-1},q)}\right)^{N-1}
	\prod_{\alpha;\text{root}} \frac{\theta_1(x^{\alpha}, q)}{\theta_1(y^{-1}x^{\alpha}, q)}.
\end{align}

\subsection{Whole result of the one loop determinant}\label{Subsec:whole1loop}

We consider now three chiral multiplets and one vector multiplet. The whole result is given by the product of them:
\begin{align}\label{Eq:ResultCV}
Z_{\Phi_1} Z_{\Phi_2} Z_{\Phi_3} Z_\text{vec} = 1.
\end{align}
Then, the these contribution is trivial. The remaining integral of the partition function is
\begin{align}\label{Eq:result}
Z_{\mathbb{T}^2\times \Sigma_\bold{g}} 
	&= \prod_{x^2,x^3\in \Sigma_\bold{g}} \int du \int_{\mathcal{M}} \mathcal{D}\Phi_0.
\end{align}
The first integral in eq.\eqref{Eq:result} is the integral of the holonomy defined in eq.\eqref{Eq:Fixedpt}.
The second integral is over the moduli space defined by the localized condition \eqref{Eq:Fixedpt}. $\mathcal{D}\Phi_0$ is the measure on it.
We performed the integral on the flat space part. Therefore, the dependence of the internal space, that is the Riemann surface, is expressed as the product in eq.\eqref{Eq:result}.

\section*{Acknowledgement}

I would like to thank Shoichi Kawamoto, Wen-Yu Wen, Satoshi Yamaguchi, Masahito Yamazaki and Yutaka Yoshida for useful discussion and comments. 

\appendix
\section{Eta and theta functions}\label{Sec:EtaTheta}
We defined the variables $q:=e^{2\pi i \tau}$ and $y:=e^{2\pi i z}$.

Eta function is defined as
\begin{equation}\label{Def:Eta}
\eta(q) := q^{1/24} \prod_{n=1}^{\infty} (1-q^n).
\end{equation}

The definition of the theta functions is
\begin{align}
\theta_1 (z|\tau) &:= -i\sum_{r\in \mathbb{Z}+1/2} 	(-1)^{r-1/2} y^r q^{r^2/2}.
\end{align}
Using infinite product these function can also be expressed as
\begin{align}\label{Def:Theta2}
\theta_1 (z|\tau) &= -iy^{1/2}q^{1/8} \prod_{n=1}^{\infty} (1-q^n)(1-yq^n)(1-\frac{1}{y}q^{n-1}).
\end{align}
In this paper we use the notation $\theta_1(z|\tau) =: \theta_1(y, q)$ and Eta function \eqref{Def:Eta} is written denotes as $\eta(q)$.
We use the useful formula which relates theta and eta functions:
\begin{align}\label{Eq:ThetaEta}
\lim_{\xi\rightarrow 1}\frac{1}{1-\xi} \theta_1(\xi,q) = i\eta(q)^3.
\end{align}
Let us proof it by using the expression eq.\eqref{Def:Theta2}, 
\begin{align}
\text{LHS of } \eqref{Eq:ThetaEta} 
&= \lim_{\xi\rightarrow 1} \frac{-i \xi^{1/2}q^{1/8} }{1-\xi}
	(1-\frac{1}{\xi})\prod_{n=1}^{\infty}  (1-q^n)(1-\xi q^{n})(1-\frac{1}{\xi}q^{n})\nonumber\\
&= i(q^{1/8} )
	\left(\prod_{n=1}^{\infty}  (1-q^n)\right)^3
= i\eta(q)^3.
\end{align}

\providecommand{\href}[2]{#2}\begingroup\raggedright\endgroup

\end{document}